\documentstyle[aps,preprint]{revtex}
\tightenlines

\begin{document}

\title{Green's Function Monte Carlo study  
of SU(3) lattice gauge theory in (3+1)D}

\author{C. J. Hamer \cite{email}, M. Samaras and R. J. Bursill \cite{uandm} 
}

\address{
School of Physics,
University of New South Wales,\\
Sydney, 2052, Australia\\
}

\maketitle

\begin{abstract}
	A `forward walking' Green's Function Monte Carlo algorithm is
used to obtain expectation values for SU(3) lattice Yang-Mills theory in
(3+1) dimensions. The ground state energy and Wilson loops are
calculated, and the finite-size scaling behaviour is explored. Crude
estimates of the string tension are derived, which agree with previous
results at intermediate couplings; but more accurate results for larger
loops will be required to establish scaling behaviour at weak coupling.
\end{abstract}

\section{Introduction}
\label{sec:intro}

	Classical Monte Carlo simulations provide a very powerful and
accurate method for the study of Euclidean lattice gauge theories. In
the Hamiltonian formulation \cite{kogut}, on the other hand,
 the corresponding
quantum Monte Carlo methods have been somewhat neglected. Here we
present a study of SU(3) Yang-Mills theory in (3+1) dimensions, using
the Green's Function Monte Carlo approach \cite{kalos}, adapted to 
lattice gauge theory by Chin et al \cite{chin1}.

	Quantum Monte Carlo methods in Hamiltonian lattice gauge theory
have a somewhat chequered history.
The first calculations used a
strong-coupling basis involving discrete ``electric field'' link
variables, and a ``Projector Monte Carlo'' approach 
\cite{blank,degrand}, which used the Hamiltonian itself to project out 
the ground state. A later version of this was the ``stochastic 
truncation'' approach of Allton et al \cite{allton}. Using this 
approach one can successfully compute string tensions and mass gaps for 
Abelian models \cite{hamer1}. For non-Abelian models, however, some
 technical problems arose
\cite{hamer1}. The use of a Robson-Webber 
recoupling scheme \cite{robson} at the 
lattice vertices requires the use of Clebsch-Gordan coefficients or 
6$j$-symbols, which are not known to high order for $SU(3)$; and 
furthermore, the `minus sign' problem rears its head, in that 
destructive interference occurs between different paths to the same 
final state. It may well be that a better choice of strong-coupling 
basis, such as the `loop representation', might avoid these problems; 
but this has not yet been demonstrated.

	In the meantime, Heys and Stump \cite{heys1} and Chin et al.
\cite{chin1} pioneered the 
use of ``Green's Function Monte Carlo'' (GFMC) or ``Diffusion Monte
Carlo'' techniques in Hamiltonian LGT, in conjunction with a
weak-coupling representation involving continuous gauge field link
variables. This was successfully adapted to non-Abelian Yang-Mills
theories \cite{heys2,chin2,heyschin3,chin3}, with no minus sign 
problem arising. In this
representation, however, one is simulating the wave function in gauge
field configuration space by a discrete ensemble or density of random
walkers: it is not possible to determine the derivatives of the gauge
fields for each configuration, or to enforce Gauss's law explicitly, and
the ensemble always relaxes back to the ground state sector. Hence one
cannot compute the string tensions and mass gaps directly as Hamiltonian
eigenvalues corresponding to ground states in different sectors, as one
does in the strong-coupling representation. 
Instead, one is forced back to the more 
laborious approach used in Euclidean calculations: namely, to measure an 
appropriate correlation function, and estimate the mass gap as the 
inverse of the correlation length. 
We have introduced the `forward-walking' technique, well-known in
many-body theory \cite{kalos,kalos1,liu,whitlock}, to measure 
the expectation values and
correlation functions. The technique has been demonstrated for the cases
of the transverse Ising model in (1+1)D \cite{samaras}, and the U(1) LGT
in (2+1)D \cite{hamer3}.

	Here we apply the technique for the first time to a non-Abelian
model, namely SU(3) Yang-Mills theory in (3+1)D. The ground state energy
and Wilson loop values are calculated, and approximate values are
extracted for the string tension in the weak-coupling regime.
Comparisons are made with earlier calculations, where they are
available \cite{hamer4}.

	Our conclusions are that the method is a viable one, but
requires the use of an improved ``guiding wave function" to achieve
better accuracy. There are certain drawbacks intrinsic to the
method \cite{kalos,ceperley}, such as the necessity to use a
 branching algorithm and a
guiding wave function, which tend to introduce substantial errors into
the results, both statistical and systematic.
For these reasons, it may be preferable to employ a Path Integral Monte
Carlo approach to these models, which avoids the problems above. 

	Our methods are presented in Section II, the results are
outlined in Section III, and our conclusions are discussed in Section
IV.

\section{METHOD}
\label{sec2}

\subsection{Lattice Hamiltonian}

	The Green's Function Monte Carlo formalism has been adapted to
SU(2) Yang-Mills theory by Chin, van Roosmalen, Umland and
Koonin \cite{chin2}, and sketched for the SU(3) case by Chin, Long and
Robson \cite{chin3}. Here we provide a slightly fuller discussion of the
SU(3) case, following the earlier treatment of Chin, van Roosmalen et
al \cite{chin2}.

The SU(3) lattice Hamiltonian is given by \cite{chin3} 
\begin{equation}
H = \frac{g^{2}}{2a}\{ \sum_{l} E^{a}_{l}E^{a}_{l} -
\frac{\lambda}{3}\sum_{p}Tr(U_{p} + U_{p}^{\dag})\} \label{1}
\end{equation}
where $E_{l}^{a}$ is a component of the electric field at link l,
$\lambda = 6/g^{4}$, the index $a$ runs over the 8 generators of SU(3),
and $U_{p}$ denotes the product of four link operators around an
elementary plaquette.

Commutation relations between electric field and link operators are
\begin{equation}
[E_{l}^{a},U_{l'}] = \frac{1}{2}\lambda^{a}U_{l}\delta_{ll'}, \label{2}
\end{equation}
choosing the $E_{l}^{a}$ as left generators of SU(3), where the
$\{\lambda^{a}, a = 1,8\}$ are the Gell-Mann matrices for SU(3). We will
work with the dimensionless operator
\begin{equation}
H = \frac{1}{2}\sum_{l}E_{l}^{a}E_{l}^{a} -\frac{\lambda}{6} \sum_{p}
Tr(U_{p} + U_{p}^{\dag}) \label{3}
\end{equation}

	The link variables are elements of the group SU(3) in the
fundamental representation
\begin{equation}
U = \exp(-i\frac{1}{2}\lambda^{a}A^{a})
\end{equation}
There is no simple equivalent of the quaternion representation for SU(2).
Following Beg and Ruegg \cite{beg}, we can represent
\begin{eqnarray}
U & = & 
 \left( \begin{array}{ccc}
z_{1} & z_{2} & z_{3} \\
u_{1} & u_{2} & u_{3} \\
w_{1} & w_{2} & w_{3} 
\end{array} \right) 
\end{eqnarray}
where {\bf z,u,w} are three-dimensional complex vectors; then if U is to
be unitary, we require {\bf z,u,w} to be orthonormal; and if
U is to have determinant unity, we require
\begin{equation}
\epsilon_{ijk}z_{i}u_{j}w_{k} = 1
\end{equation}
which is satisfied if
\begin{equation}
w_{i} = \epsilon_{ijk}z_{i}^{*}u_{k}^{*}
\end{equation}

	One possible representation which satisfies these conditions
is
\begin{eqnarray}
z_{1} & = & (1-i(x_{3} +x_{8}/\surd{3}))/N_{1}, \nonumber \\
z_{2} & = & (-x_{2} - i x_{1})/N_{1}, \nonumber \\
z_{3} & = & (-x_{5}-ix_{4})/N_{1} 
\end{eqnarray}
where
\begin{equation}
N_{1} = [1 + (x_{3} + x_{8}/\surd{3})^{2} + x_{1}^{2} + x_{2}^{2} +
x_{4}^{2} + x_{5}^{2}]^{1/2};
\end{equation}
and
\begin{equation}
u_{3}  =  (-x_{7}-ix_{6})/N_{2}  \equiv  \tilde{u_{3}}/N_{2}, \nonumber 
\end{equation}
\begin{equation}
u_{2}  =  (1-i(x_{8}/\surd{3} - x_{3}))/N_{1}  \equiv 
\tilde{u_{2}}/N_{2}, \nonumber 
\end{equation}
\begin{equation}
u_{1}  =  \tilde{u_{1}}/n_{2}, \hspace{3mm}  \tilde{u_{1}}  =  -[
\tilde{u_{2}}z_{2}^{*} + \tilde{u_{3}}z_{3}^{*}]/z_{1}^{*}
\end{equation}
where
\begin{equation}
N_{2} = [|\tilde{u_{1}}|^{2} + |\tilde{u_{2}}|^{2} +
|\tilde{u_{3}}|^{2}]^{1/2}
\end{equation}
and
\begin{equation}
w_{i} = \epsilon_{ijk}z_{j}^{*}u_{k}^{*}.
\end{equation}
This involves 8 unrestricted parameters $\{x_{a}, a = 1 .. 8\}$, as
expected. For small {\bf x},
\begin{equation}
U \approx I - i\lambda^{a}x^{a}
\end{equation}
i.e.
\begin{equation}
 x^{a} \approx A^{a}/2,
\end{equation}
where $A^{a}$ is the gauge field on that link.
	The product of two link variables can be found by simple
matrix multiplication.

\subsection{Green's Function Monte Carlo method}

	The Green's Function Monte Carlo method employs the operator
$\exp(-\tau(H-E))$, i.e. the time evolution operator in imaginary time,
as a {\it projector} onto the ground state $|\psi_{0} \rangle$:
\begin{eqnarray}
 |\psi_{0} \rangle & \propto & \lim_{\tau \rightarrow
\infty}\{e^{-\tau(H-E)}|\Phi \rangle \} \nonumber\\ 
 & = & \lim_{\Delta\tau \rightarrow 0, N\Delta\tau \rightarrow \infty}
e^{-N\Delta\tau(H-E)}|\Phi\rangle
\label{4}
\end{eqnarray}
where $|\Phi \rangle$ is any suitable trial state. To procure some
variational guidance, one performs a ``similarity transformation" with
the trial wave function $\Phi$, and evolves the {\it product}
$\Phi|\psi_{0}\rangle$ in imaginary time. The heart of the procedure is
the calculation of the matrix element corrersponding to a single small
time step $\Delta\tau$. Chin et al \cite{chin2} show that 
\begin{eqnarray}
\langle {\bf x}'| \Phi e^{-\Delta\tau(H-E)}\Phi^{-1}|{\bf x}\rangle & =
& \prod_{l}\langle
U_{l}'|N\{ \exp (-\frac {1}{2} \Delta \tau E_{l}^{a}E_{l}^{a}) 
\exp[\Delta\tau E^{a}_{l}
(E^{a}_{l}\ln\Phi)]\}|U_{l}\rangle \nonumber\\
 & & \exp\{\Delta\tau [E-\Phi^{-1}H\Phi({\bf
x})]\} + O(\Delta\tau^{2}) \nonumber\\
 & \equiv & p({\bf x',x})w({\bf x}) + O(\Delta\tau^{2}) \label{eq5}
\end{eqnarray}
where ${\bf x} = \{U_{l}\}$ denotes an entire lattice configuration of
link fields.

	The product $\Phi|\psi\rangle$ is simulated by the density of an
ensemble of random walkers, as in the SU(2) case. At the kth. step, the
`weight' of each walker at ${\bf x}_{k}$ is multiplied by $w({\bf
x}_{k})$ and the next ensemble $\{{\bf x}_{k+1}\}$ is evolved from
$\{{\bf x}_{k}\}$ according to the matrix element $p({\bf
x_{k+1},x_{k}})$. The effect of $p({\bf x_{k+1},x_{k}})$ is to alter
each link variable $U_{l}$ in $\{{\bf x_{k}}\}$ to $U_{l}'$ by a
Gaussian random walk plus a ``drift step" guided by the trial wave
function:
\begin{equation}
U' = \Delta U U_{d} U
\end{equation}
where
$U_{d} = \exp [i\frac{1}{2}\lambda^{a}(i \Delta \tau E^{a} \ln \Phi )]$
is the drift step, and $\Delta U$ is an SU(3) group element randomly 
chosen from a
Gaussian distribution around the identity, with variance
$\langle\Delta s^{2}\rangle = 8\Delta\tau$ (i.e. $\Delta \tau$ for each
index {\it a}), where
\begin{equation}
\langle \Delta s^{2}\rangle \approx \sum_{a}A^{a}A^{a} = 8\Delta \tau,
\hspace{5mm} $(small $A^{a})$$
\end{equation}
or
\begin{equation}
\langle x^{a}x^{a}\rangle = \frac{1}{4}\langle A^{a}A^{a}\rangle \approx
\frac{\Delta \tau}{4}, \hspace{5mm} $each$ \hspace{1mm}a.
\end{equation}

	The simulation is carried out for a large number of iterations
$\Delta\tau$, until an equilibrium distribution $\Phi|\psi_{0}\rangle$ is
reached. The energy E in (\ref{eq5}) is adjusted after each iteration so as to
maintain the total ensemble weight constant. The average value of E can
then be taken as an estimate of $E_{0}$, the ground-state energy.

	As time evolves, the weights of some walkers grow larger, while
others grow smaller, which would produce an increased statistical error.
To avoid this, a ``branching" process is employed, whereby a walker with
weight larger than some threshold is split into two independent walkers,
while others with weights lower than another threshold are amalgamated. We
use Runge's technique \cite{runge} for this purpose.

\subsection{Trial Wave Function}

	The trial wave function is chosen to be the one-parameter
form \cite{chin3}
\begin{equation}
\Phi = \exp[\alpha\sum_{p}Tr(U_{p} + U_{p}^{\dag})]
\end{equation}

	Then the drift step is \cite{chin2}
\begin{eqnarray}
U_{d} & = & \exp[i\frac{\lambda^{a}}{2}(i\Delta \tau E^{a}\ln\Phi)]
\nonumber \\
 & \equiv & \exp[-i \frac{\lambda^{a}}{2}A^{a}_{l}]
\end{eqnarray}
for each link, where
\begin{equation}
A^{a}_{l} = -i \Delta \tau \frac{\alpha}{2}\sum_{p \in l}
Tr[\lambda^{a}U_{l} .. U_{4}^{\dag} - h.c.]
\end{equation}
i.e.
\begin{equation}
x_{l}^{a} \approx \frac{A_{l}^{a}}{2} =
-i\frac{\alpha\Delta\tau}{4}\sum_{p \in l} Tr[\lambda^{a}U_{l} ..
U_{4}^{\dag} - h.c.]
\end{equation}
(note: the effect of $E^{a}_{l}$ on a plaquette operator is to `insert'
a $\lambda^{a}/2$ in front of the appropriate link operator $U_{l}$, to
be followed by the remaining link operators in the plaquette, taken in
the direction of the link l).

	Finally, the trial energy factor is
\begin{eqnarray}
\Phi^{-1}H\Phi & = & \sum_{l}\{\frac{\alpha^{2}}{8}(\sum_{p \in l}
Tr[\lambda^{a}U_{l} .. U_{4}^{\dag} - h.c.])^{2} 
+(\frac{2\alpha}{3} - \frac{\lambda}{24})\sum_{p \in l} Tr(U_{p} +
U_{p}^{\dag})\} .
\end{eqnarray}

	Therefore the weight factor is
\begin{eqnarray}
w({\bf x}) & = & \exp\{\Delta\tau(E - \Phi^{-1}H\Phi)\} \nonumber \\
 & = & \exp\{\Delta\tau(E_{trial} - (\frac{2\alpha}{3} -
\frac{\lambda}{24})\sum_{l}\sum_{p \in l} 2Re\{Tr U_{p}\} \nonumber \\
 & & + \frac{\alpha^{2}}{8}\sum_{l}(\sum_{p \in l}2Im\{Tr[\lambda^{a}U_{l} 
.. U_{4}^{\dag}]\})^{2}))\} .
\end{eqnarray}

\subsection{Forward Walking estimates}
\label{subsec: for}

	The ``forward walking" technique is used to estimate expectation
values \cite{kalos}. Its application to the U(1) lattice gauge theory in
(2+1)D was discussed by Hamer et al \cite{hamer3}. It is based on the
following equation: for an operator Q,
\begin{eqnarray}
<Q>_{0} & = & \frac{ \langle \psi_{0} |Q| \psi_{0} \rangle }
                   { \langle \psi_{0}| \psi_{0} \rangle } 
\nonumber\\
        &
\stackrel{\textstyle \sim}{\scriptscriptstyle J \rightarrow \infty}
           &
\frac{ \langle \Phi\ | K^{J} Q | \psi_{0} \rangle }
{ \rangle \Phi | K^{J} | \psi_{0} \rangle }
\nonumber\\
       & = &
\frac{ \sum \tilde{K}({\bf x}_{J}, {\bf x}_{J-1}) \ldots 
\tilde{K}({\bf x}_{2}, {\bf x}_{1})
Q({\bf x}_{1}) \tilde{\psi}_{0}({\bf x}_{1})
     }{
\sum \tilde{K}({\bf x}_{J}, {\bf x}_{J-1}) \ldots
\tilde{K}({\bf x}_{2}, {\bf x}_{1}) \tilde{\psi}_{0}({\bf x}_{1})
      }
\label{eq16}
\end{eqnarray}
where $K({\bf x}_{J},{\bf x}_{J-1})$ is the evolution operator for
time $\Delta \tau$, and $\tilde{K}({\bf x}_{J},{\bf x}_{J-1})$ is 
the same operator in the similarity transformed basis. Again we
 have assumed that the operator $Q$ is diagonal in the basis of 
plaquette variables ${\bf x}$.

This equation is implemented by \cite{kalos1,liu,whitlock}
the following procedure:
\begin{itemize}
\item[i)]
Record the value $Q({\bf x}_{i})$ for each ``ancestor" walker at the
 beginning of a measurement;
\item[ii)]
Propagate the ensemble as normal for $J$ iterations, keeping a record
 of the ``ancestor" of each walker in the current population;
\item[iii)]
Take the weighted average of the $Q({\bf x}_{i})$ with respect to the 
weights of the descendants of ${\bf x}_{i}$ after the $J$ iterations, 
using sufficient iterations $J$ that the estimate reaches a `plateau'.
\end{itemize}

\section{Results}

	Simulations were carried out for LxLxL lattices up to L=8 sites,
using runs of typically 4000 iterations and an ensemble size of 250
to 1000 depending (inversely) on lattice size. These statistics are
approximately 100 times less than those used in the U(1)
calculations \cite{hamer3}, but about 100 times greater than those
used in the previous SU(3) calculations of Chin et al \cite{chin3}. Time
steps $\Delta\tau$ of 0.01 and 0.05 ``seconds" were used, with each 
iteration consisting of 5 sweeps and 1 sweep through the lattice, 
respectively,followed by a branching process.
The first 400 iterations were discarded to allow for equilibration.
The data were block averaged over blocks of up to 256 iterations, to
minimize the effect of correlations on the error estimates.

	The results taken at $\Delta\tau = 0.01$ and $ \Delta\tau =
0.05$ were extrapolated linearly to $\Delta \tau = 0$. Figure 1
demonstrates that the dependence of the ground-state energy on
$\Delta\tau$ is approximately linear.

	The variational parameter $c$ was given values as shown in Table
I. These are essentially the values used by Chin et al \cite{chin3},
obtained from a variational Monte Carlo calculation. We checked that
these were approximately the optimum values for small lattices.

	Forward-walking measurements were taken over $J$ iterations,
 where $J$ ranged from
20 to 100, depending on the coupling $\lambda$. Ten separate
measurements were taken over this time interval, in order to check
whether the value measured by forward-walking had reached equilibrium. A
new measurement was started soon after the previous one had finished.

\subsection{Ground-state Energy}

	The dependence of the ground-state energy on the variational
parameter $c$ is illustrated in Figure 2. It can be seen that the energy
reaches a broad minimum at about the expected value ($c = 0.33$ at this
coupling).

	Our estimates of the ground-state energy are listed in Table I,
as a function of the coupling $\lambda$ and lattice size $L$. The
dependence on lattice size is illustrated in Figure 3, at two fixed
couplings $\lambda = 3.0$ and $\lambda = 5.0$. In the
``strong-coupling" case, $\lambda = 3.0$, it can be seen that the
results converge exponentially fast in $L$, whereas in
the ``weak-coupling" regime, $\lambda = 5.0$, the convergence is more
like $1/L^{4}$ at these lattice sizes. This behaviour merits some
further explanation. 

	A similar phenomenon occurs in the case of the U(1) theory 
in (2+1)D \cite{hamer5,hamer1,hamer3}. In the
strong-coupling regime, where the mass gap is large, the usual 
exponential convergence occurs. In
the weak-coupling regime, however, where the mass gap M is very small,
 the finite-size scaling
behaviour for small lattice sizes is that of a massless theory, and 
it is only at much larger
lattice sizes $ L \approx 1/M$ that a crossover to exponential 
convergence occurs. In the U(1)
case, it has been shown \cite{hamer1,hamer3} that the finite-size 
scaling behaviour at small $L$ is well described
by an ``effective Lagrangian" approach, using the Lagrangian for free,
 massless photons that the
model was originally constructed to simulate. In the same way, 
an ``effective Lagrangian"
corresponding to free, massless gluons (non-interacting QCD) should 
describe the finite-size
behaviour in the present case, in line with the idea of asymptotic
 freedom. By analogy with the
(2+1)D case, we expect a $1/L^{4}$ dependence for the corrections 
to the ground-state energy per
site. We hope to pursue this analysis further at a later date.

	An anomalous feature in Figure 3b)  is that the $L = 8$
 point lies well out of line
with the others. This occurs at other couplings also.
We suspect that the results for $L = 8$ are not reliable, and 
that the trial wave function will
have to be further improved to give reliable results for such
 large lattices. Supporting evidence
for this will be presented below.

We have made estimates of the bulk limit, extrapolating mainly from the
smaller L values where possible, and the results are listed in Table
II. Our present estimates generally lie a little below those of Chin et
al \cite{chin3}, and we believe them to be more accurate in view of our
greater statistics. 
The estimates for the bulk ground-state energy per site are graphed
as a function of coupling in Figure 4, where they are compared with 
previous estimates \cite{hamer4} obtained by an `Exact Linked Cluster 
Expansion' (ELCE) procedure, and with the
asymptotic weak-coupling series \cite{hofsass} 

\begin{equation}
\epsilon_{0} \sim -3\lambda + 7.798 \lambda^{1/2}, \hspace{3mm} \lambda
\rightarrow \infty.
\end{equation}
The Monte Carlo results agree very well with the ELCE estimates, and
appear to match nicely onto the expected weak-coupling behaviour for
$\lambda \geq 6$.

\subsection{Wilson Loops}

The forward-walking method was used to estimate values for the $m x n$
Wilson loops, $W(m,n)$. 
Figure 5 shows an example, namely the estimate of the mean plaquette
$W(1,1)$ as a function of $J$ for the case $ L=6$, $\lambda =1.5$. 
It can be seen
that the estimate relaxes exponentially towards a plateau value as the
number of iterations $J$ increases: an exponential fit is performed to
estimate the asymptotic value. It can also be seen that the statistical
error on each point is much larger than the point-to-point variation:
the estimates are highly correlated, and all the points tend to move up
and down together as one goes from one sample to the next.

A problem that arises in these measurements is the loss in
statistical accuracy at large couplings. At large couplings the
weights of the random walkers vary rapidly with time, and it can easily
happen that during a measurement the descendants of all the 
`ancestor' walkers but one die
out from the ensemble, at which point the result `freezes', and the
number of members of the ensemble has effectively been reduced to one.
This inevitably means a severe loss of statistical accuracy.

A graph of the `mean plaquette' $W(1,1)$ versus the variational
parameter $c$ is shown in Figure 6. Another problem is immediately
apparent. The estimate for $W(1,1)$ is not independent of $c$, in fact 
it depends linearly on $c$ over this range, 
and the size of the variation is
such that the probable systematic error due to the choice of $c$ is an
order of magnitude larger than the random statistical error in the
results. Thus it would be advatageous in future studies to put more
effort into improving the trial wave function, rather than merely
improving the statistics.

Figure 7 shows examples of the dependence of the results on lattice size
$L$. Once again, the results at strong coupling $\lambda = 1.5$ converge
exponentially fast, while those at the weak coupling value $\lambda =
5.0$ can be approximately fitted by a $1/L^{4}$ dependence. The $L = 8$
value and even the $L = 6$ value again lie off the trend of the smaller
lattices, and are probably not very reliable.

Estimates of the bulk limit
 are listed in Table
II. The estimates for the mean plaquette are graphed as a function of
coupling $\lambda$ in Figure 8, and compared with series estimates at
strong and weak coupling \cite{hamer4,hofsass}. The agreement is quite
good.

\subsection{String Tension}

	Having obtained estimates for the Wilson loop values on the bulk
lattice, one can extract estimates for the `spacelike' string tension
using the Creutz ratios:
\begin{equation}
Ka^{2} \simeq R_{n} = - \ln \left[\frac{W(n,n)W(n-1,n-1)}{W(n,n-1)^{2}}
\right]
\end{equation}

or the cruder 2-point estimates
\begin{equation}
R'_{n} = -\frac{1}{n}\ln\left[\frac{W(n,n)}{W(n,n-1)} \right]
\end{equation}

	The results are shown in Figure 9. Also shown in Figure 9 are
some previous estimates derived from the `axial' string tension,
obtained \cite{hamer4} 
  using an
`Exact Linked Cluster Expansion' (ELCE) method. 
The axial string tension $aT$ is calculated as an energy per link, and
must be converted to a dimensionless, `spacelike' tension by dividing by
the `speed of light' c,
\begin{equation}
Ka^{2} = \frac{aT}{c},
\end{equation}
where \cite{hasenfratz}
\begin{equation}
c \sim \frac{2}{g^{2}}[1 - 0.1671g^{2}] = \sqrt{\frac{2\lambda}{3}}
\left[1-0.1671\sqrt{\frac{6}{\lambda}}\hspace{2mm}\right],
 \hspace{3mm} \lambda \rightarrow \infty.
\end{equation}
We have also used the weak-coupling relationship between the scales of
Euclidean and Hamiltonian lattice Yang-Mills theory calculated by
Hasenfratz et al \cite{hasenfratz} to plot the results against the 
Euclidean coupling $\beta = 6/g^{2}_{E}$, where
\begin{equation}
\beta = \sqrt{6\lambda} -0.01308
\end{equation}

	It can be seen that the present GFMC results are in rough
agreement with the axial string tension results in the region $4 \leq
\beta \leq 5$, which is also the region where
 the `roughening' transition
occurs in the string tension \cite{hamer4}. For $\beta > 5$, however, the
Creutz ratio $R_{2}$ runs above the ELCE estimate, and shows no sign of
the expected crossover to an exponentially decreasing scaling behaviour
at $\beta \simeq 6$. We presume that this is a finite-size effect, and that
the Creutz ratios $R_{n}$ for larger $n$ will show a substantial decrease
in the `weak-coupling' regime $\beta \geq 6$. That is certainly the
pattern seen in the Euclidean calculations \cite{creutz}, or in the
U$(1)_{2+1}$ model \cite{hamer3}. Unfortunately, however, our present 
results for the larger Wilson loops are not of sufficient accuracy to 
allow worthwhile estimates of $R_{n}$ for $n \geq 2$.

\section{Summary and Conclusions}

	We have presented the results of a new Green's Function Monte
Carlo study of the SU(3) Yang-Mills theory in the (3+1)D Hamiltonian
formulation. A forward-walking method has been used to estimate values
for the Wilson loops as well as the ground-state energy, and hence some
rather crude estimates of the string tension have been extracted.
Comparisons have been made with an earlier Hamiltonian calculation of the
axial string tension \cite{hamer4}. The two sets of results agree in the
`roughening' region; but our Monte Carlo results do not extend to the
large Wilson loops that would be required to demonstrate `scaling'
behaviour in the weak-coupling regime.

	Some significant problems with the GFMC method have emerged from
this study. The `forward-walking' technique was introduced specifically
to avoid any variational bias from the trial wave function
\cite{kalos,kalos1,liu,whitlock}. As it
turns out, however, the results for the Wilson loops show a substantial
dependence on the trial wave function parameter $c$.
 The systematic error due to this
dependence is an order of magnitude larger than the statistical error,
so it would pay to put more effort in future studies into improving the
trial wave function, rather than simply increasing the statistics.
Furthermore, the effective ensemble size decreases during each
measurement as the descendants of each `ancestor' state die out, and
this produces a substantial loss in statistical accuracy at weak
coupling, as well.

	It would be preferable if one were able to do away entirely with
all the paraphernalia of trial wave function, weights, branching
algorithms, etc, and just rely on some sort of Metropolis-style
accept/reject algorithm to produce a correct distribution of walkers.
Within a quantum Hamiltonian framework, a way is known to do this,
namely the Path Integral Monte Carlo (PIMC) approach \cite{ceperley}.
 We conclude
that the PIMC approach may be better suited than GFMC to the study of
large and complicated lattice Hamiltonian systems. 

\section*{Acknowledgments} This work is supported by the Australian 
Research Council. Calculations were performed on the SGI Power Challenge 
Facility at the New South Wales Centre for Parallel Computing and the 
Fujitsu VPP300 vector machine at the Australian National Universtiy 
Supercomputing Facility: we are grateful for the use of these facilities.

\newpage
\squeezetable
\begin{table}[p]
\begin{center}
\begin{tabular}{|c|c|c|c|c|c|c|}  \hline
$\lambda$ & 1.5 & 3.0 & 4.0 & 5.0 & 6.0 & 9.0 \\  \hline\hline
c   & 0.10 & 0.20 & 0.27 & 0.33 & 0.40 & 0.58 \\ \hline
 \multicolumn{7}{l}{Ground state energy per site, $\epsilon_o$}    \\
\hline
L = 2 & -0.15449(2)  & -0.68081(1) & -1.3212(5) & -2.359(1) & -3.666(2)
&
-8.389(6) \\
3 & -0.15438(4) & -0/6764(2) & -1.2846(3) & -2.187(1)  & -3.431(2) &
-8.042(4)  \\
4 & -0.15436(1) & -0.6764(1) & -1.2836(8) & -2.160(3) & -3.322(4) &
-7.94(1)   \\
6 & -0.15436(2) & -0.6765(1) & -1.2809(7) & -2.133(2) & -3.234(3) &
-7.613(9) \\
8 &   - & -0.6759(2) & -1.2705(4) & -2.094(2) & -3.154(2)  & -7.402(6)
\\ \hline
 \multicolumn{7}{l}{ Wilson loops, W(1,1)}                \\ \hline
L = 2 & 0.0721(2) & 0.1671(3) & 0.273(2) & 0.396(6) & 0.460(7) &
0.576(4)
\\
3 & 0.0718(2) & 0.1641(4) & 0.245(2) & 0.346(4) &  0.437(5) & 0.562(2)
\\
4 & 0.0718(2) &  0.1647(7) & 0.240(2) & 0.330(4) & 0.401(5) &
0.556(4) \\
6 & 0.0718(1) & 0.1648(4) & 0.240(3) & 0.310(1) & 0.383(2) & 0.550(3)
\\
8 &  - & 0.1633(8) & 0.228(1) & 0.296(1) & 0.365(2) & 0.545(2) \\
\hline
 \multicolumn{7}{l}{ W(2,1)}                              \\ \hline
3 & 0.0059(1) & 0.0318(3) & 0.070(1) & 0.146(4) & 0.234(6) & 0.376(3)
\\
4 & 0.0061(1) & 0.0322(4) & 0.067(2) & 0.130(4) & 0.182(4) & 0.355(6)
\\
6 & 0.0061(1) & 0.0322(3) & 0.069(2) & 0.109(1) & 0.165(2) & 0.329(3)
\\
8 & - & 0.0309(5) & 0.057(1) & 0.097(1) & 0.144(2) & 0.313(2)  \\
\hline
 \multicolumn{7}{l}{ W(2,2)}                              \\ \hline
4 & - & 0.0010(7) & 0.006(2) & 0.022(5) & 0.044(5) & 0.183(7) \\
6 & - & 0.0022(3) & 0.009(2) & 0.018(1) & 0.037(2) & 0.137(4) \\
8 & - & 0.0011(6) & 0.005(1) & 0.013(1) &  0.023(2) &
0.111(3)  \\ \hline
 \multicolumn{7}{l}{ W(3,2)}                              \\ \hline
6 & - & - & - & 0.004(1) & 0.006(2) & 0.059(3)  \\
8 & - & - & - & 0.001(1) & 0.004(1) & 0.039(2) \\ \hline
 \multicolumn{7}{l}{ W(3,3)}\\ \hline
6 & - & - & - & - & - & 0.020(3) \\
8 & - & - & - & - & - & 0.009(2) \\ \hline
\end{tabular}
\end{center}
\caption{Variational parameter
 $c$, ground state energy per site
$\epsilon_o$ and Wilson loops $W(n,m)$ as functions of the lattice size
$L$ and coupling $\lambda$.\  }
\label{table1}
\end{table}
\begin{table}
\begin{center}
 \begin{tabular}{|c|c|c|c|c|c|c|}  \hline
 $\lambda$ & 1.5 & 3.0 & 4.0 & 5.0 & 6.0 & 9.0 \\ \hline\hline
\multicolumn{7}{l}{Ground state energy per site, $\epsilon_o$}\\ \hline
this work & -0.15436(2) & -0.6764(1) & -1.284(2) & -2.16(2) &
-3.25(2) & -7.9(1) \\
Chin et al [13] & & -0.675(0) & -1.275(3) & -2.088(6) & -3.183(6) &
-7.50(3) \\
\hline
\multicolumn{7}{l}{Wilson loops W(1,1)}\\ \hline
this work & 0.0718(2) & 0.165(1) & 0.240(1) & 0.32(1) & 0.39(1) &
0.554(3)  \\
Chin et al [13] & & 0.1605(5) &  & 0.298(5) & 0.377(3) & 0.539(3)  \\
\hline
\multicolumn{7}{l}{W(2,1)}\\ \hline
this work & 0.0061(2) & 0.0322(5) & 0.068(2) & 0.12(1) & 0.165(5) &
0.34(1) \\ \hline
\multicolumn{7}{l}{W(2,2)}\\ \hline
this work &  & 0.0021(1) & 0.008(4) & 0.018(4) & 0.035(7) & 0.13(1)
\\ \hline
\multicolumn{7}{l}{W(3,2)}\\ \hline
this work &  & &  & 0.004(2) & 0.006(2)  & 0.05(1)  \\ \hline
\multicolumn{7}{l}{W(3,3)}\\ \hline
this work &  & &  & & & 0.01(1)  \\ \hline
\end{tabular}
\end{center}
\caption{Estimates of the bulk ground-state energy per site and Wilson
loops as functions of $\lambda$.\ }
\label{table2}
\end{table}

\begin{figure}[htbp]
\caption{
Estimated ground-state energy for lattice size $L = 4$ and coupling
$ \lambda = 3.0$, $c = 0.20$, as a function of time step $\Delta \tau$.
}
\label{fig1}
\end{figure}
\begin{figure}[htbp]
\caption{
Estimated ground-state energy as a function of the variational
parameter $c$, for $L = 6, \lambda = 5.0$.
}
\label{fig2}
\end{figure}
\begin{figure}[htbp]
\caption{
Ground-state energy per site graphed against $1/L^{4}$, where L is the
lattice size: a) at coupling $\lambda = 3.0$, b) at coupling $\lambda =
5.0$. The lines are merely to guide the eye.
}
\label{fig3}
\end{figure}
\begin{figure}[htbp]
\caption{
The bulk ground-state energy per site graphed against coupling
$\lambda$. The points are our Monte Carlo estimates; the solid line
represents earlier ELCE estimates[19]; and the dashed line represents the
asymptotic weak-coupling behaviour.
}
\label{fig4}
\end{figure}
\begin{figure}[htbp]
\caption{
Measured value for the Wilson loop $W(1,1)$ as a function of the number of
forward-walking iterations J, for the case $L = 6, \lambda = 1.5$. The
solid line is an exponential fit to the data.
}
\label{fig5}
\end{figure}
\begin{figure}[htbp]
\caption{
Estimated value for the mean plaquette $W(1,1)$ as a function of the
variational parameter $c$, for $L= 6, \lambda = 5.0$.
}
\label{fig6}
\end{figure}
\begin{figure}[htbp]
\caption{
The mean plaquette $W(1,1)$ graphed against $1/L^{4}$, where L is the
lattice size: a) at coupling $\lambda = 1.5$, b) at coupling $\lambda =
5.0$. The lines are merely to guide the eye.
}
\label{fig7}
\end{figure}
\begin{figure}[htbp]
\caption{
The mean plaquette $W(1,1)$ for the bulk system graphed against coupling
$\lambda$. The solid line represents the strong-coupling series
expansion[19], and the dashed line the asymptotic weak-coupling
behaviour.
}
\label{fig8}
\end{figure}
\begin{figure}[htbp]
\caption{
The string tension $Ka^{2}$ graphed against coupling $\beta$. The
circles are obtained from ELCE estimates of the axial string 
tension[19];
the triangles are Monte Carlo estimates of $R_{2}$.
}
\label{fig9}
\end{figure}

\end{document}